\documentclass{WileyMSP-template}

\usepackage{dcolumn}
\usepackage{graphicx}
\usepackage{epstopdf}
\usepackage{mathrsfs}
\usepackage{subfigure}
\usepackage{booktabs}
\usepackage{amsmath}
\usepackage{physics}
\usepackage{dsfont}
\usepackage{amstext}
\usepackage{amssymb}
\usepackage{amsbsy}
\usepackage{bbm}
\usepackage{amsthm}
\usepackage{xcolor}
\usepackage{url}

\usepackage{cite}

\usepackage[
    colorlinks=true,
    urlcolor=blue,
    linkcolor=blue,
    citecolor=blue,
    breaklinks=true,
    hypertexnames=false,
    pageanchor=true,
    anchorcolor=black
]{hyperref}

\begin{document}
\pagestyle{fancy}
\rhead{\includegraphics[width=2.5cm]{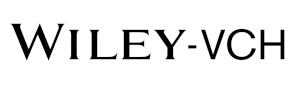}}

\title{Quantum Phase Transitions in Coherent Ising Machines:\\
 XY Model for Demonstration}

\maketitle

\author{Shuang-Quan Ma}
\author{Jing-Yi-Ran Jin}
\author{Qing Ai*}

\dedication{}

\begin{affiliations}
Shuang-Quan Ma, Jing-Yi-Ran Jin, Qing Ai*\\
School of Physics and Astronomy, Beijing Normal University,\\
and Key Laboratory of Multiscale Spin Physics (Beijing Normal University), Ministry of Education,\\
Beijing 100875, China\\
Email: aiqing@bnu.edu.cn
\end{affiliations}


\begin{abstract}
	
In this work, we study the detection of quantum phase transitions (QPTs) in coherent Ising machines (CIMs) through a spectral mapping between the one-dimensional XY spin model and a network of degenerate optical parametric oscillators (DOPOs). This exact correspondence reveals that the DOPO spectrum faithfully encodes the critical behavior of anisotropic and isotropic XY chains, while a strongly anisotropic regime provides a controlled near-transverse-field-Ising realization, with the energy gap closing precisely at the critical point. The ground-state energy density and its derivatives are analyzed, and the resulting singularities in magnetic susceptibility accurately pinpoint the quantum critical points of the corresponding spin model. These results show that CIMs not only serve as powerful platforms for solving combinatorial optimization problems but also provide a versatile tool for probing universal quantum critical phenomena, bridging quantum-spin models and photonic quantum systems.

\end{abstract}

\section{Introduction}
\label{sec:introduction}

Quantum phase transitions (QPTs) mark fundamental changes in the ground state of a quantum many-body system, driven solely by quantum fluctuations at zero temperature. Unlike classical phase transitions, which arise from thermal fluctuations, QPTs occur when a control parameter (e.g., magnetic field or coupling strength) in the systems Hamiltonian is tuned across a critical point, leading to non-analytic behavior in the ground-state energy. This abrupt change reflects a reorganization of quantum correlations and often accompanies the closing of the energy gap in the thermodynamic limit~\cite{Sachdev2011,prl2015Hwang}. The study of QPTs does not only illuminate the deep structure of quantum matter~\cite{prb2023hu,prl2020zhu,prl2022Myerson,prb2017Arya,sr2013Zhang,prx2021Hertkorn} but also bridges seemingly disparate fields such as condensed matter physics~\cite{science2022Maksimovic,prb2022wang,sr2015Bhattacharyya,prb2023Vijayan,prb2023Paviglianiti}, quantum information science~\cite{pra2018zhang,prl2019sahu,prl2018heyl,pra2008Ai}, and critical phenomena~\cite{prb2024Brillaux,prx2025Wu,np2018Kim,fp2023He}. In particular, the enhanced quantum fluctuations and long-range entanglement near critical points offer novel mechanisms for quantum sensing~\cite{pra2023tang,pra2022lv,pra2023shao,prl2019Heugel,prr2021Guan}, topological materials~\cite{prm2024Shafiei,prb2022sun,prl2024mogi}, and the engineering of exotic phases of matter~\cite{prb2022Kalinowski,pre2022Yang,nature2025Manovitz}. As such, QPTs serve as both a theoretical cornerstone and a practical frontier in the exploration of strongly correlated quantum systems.

Building upon the theoretical framework of QPTs, the one-dimensional XY model provides an analytically accessible yet physically rich platform for exploring quantum critical phenomena~\cite{prl2006Zhu,pra2022fu,pra2014sun,pra2011ma,prl2024Grazi,prl2022Nandi,prl1995Pazmandi,prb2020Zamani}. The model captures essential features of quantum magnetism, such as competition between exchange interactions and external fields, while remaining exactly solvable. This solvability therefore enables the precise identification of quantum critical points, which are characterized by the closing of the energy gap and the emergence of universal scaling behavior. This universality likewise makes it a useful model for investigating quantum battery~\cite{pra2022Ghosh,pra2021Ghosh} and quantum discord~\cite{sr2023Satoori,ap2015Liu,prb2024Ribeiro}.

Experimentally, the quantum critical phenomena of the XY model can be accessed through the coherent Ising machine (CIM), which bridges its theoretical description with a tunable physical implementation. At its core, a CIM employs degenerate optical parametric oscillators (DOPOs) to simulate qubits, enabling efficient computation of the ground state. When the pump intensity surpasses the bifurcation threshold, the system undergoes a collective spontaneous symmetry breaking, driving it from the normal phase to the superradiant phase~\cite{pra2013wang,pra2020Inui}. Originally harnessed for solving Ising-based optimization problems~\cite{sr2016Takata,science2016Inagaki,science2016McMahon,ol2023Mwamsojo,sa2025Takesue}, the nonlinear and collective dynamics of DOPO networks have since established CIMs as versatile simulators for quantum many-body phenomena, including phase transitions \cite{pra2023inaba,sa2021honjo,pra2023Takesue,prx2024Yamamura}. Beyond physics simulation, they are also used in many fields such as chemical synthesis~\cite{prr2024Mizuno,acr2021Terayama} and biology \cite{Kell2012}. Recent experimental progress has also strengthened the practical basis for scalable and stable CIM. In particular, Wei \textit{et al.} demonstrated a femtosecond-pumped CIM platform with stable operation over eight hours while solving a 100-vertex M\"obius-ladder instance~\cite{Wei2026Light}. Such advances provide an important experimental foundation for future tests of the tunable DOPO-network protocol proposed here.

Following this introduction, the paper is organized as follows. In Section~\ref{sec:model}, we introduce the XY spin model and the DOPO network, and establish the spectral mapping between them. In Section~\ref{sec:results and discussion}, we present numerical results for the ground-state energy density and its derivatives, and discuss the simulated quantum critical phenomena. We conclude in Section~\ref{sec:conclusion} and provide an outlook on future directions.

\section{Model}
\label{sec:model}

\subsection{ XY Model}
The XY model represents a fundamental quantum spin system that exhibits rich critical behavior \cite{Sachdev2011}. With $\hbar=1$, the Hamiltonian for this model is given by
\begin{equation}
H_{\text{XY}}=-\sum_{i=1}^{N}\left(J_{x}\sigma_{i}^{x}\sigma_{i+1}^{x}+J_{y}\sigma_{i}^{y}\sigma_{i+1}^{y}\right)-h\sum_{i=1}^{N}\sigma_{i}^{z},
\end{equation}
where $J_x$ ($J_y$) denote the coupling strength along the $x$ ($y$) direction, $h$ represents the transverse field, and $\sigma_i^\alpha$ ($\alpha = x,y,z$) are the Pauli matrices.

To analyze this system, we first employ the Jordan-Wigner transformation \cite{Sachdev2011} to map the spin operators to the fermionic operators as
\begin{equation}
\begin{aligned}
\sigma_{i}^{+} &= c_{i}^{\dagger}e^{i\pi\sum_{j=1}^{i-1}c_{j}^{\dagger}c_{j}}, \\
\sigma_{i}^{-} &= c_{i}e^{-i\pi\sum_{j=1}^{i-1}c_{j}^{\dagger}c_{j}}, \\
\sigma_{i}^{z} &= 2c_{i}^{\dagger}c_{i}-1.
\end{aligned}
\end{equation}
Here $\sigma_{i}^+$ and $\sigma_{i}^-$ denote the raising and lowering operators for the $i$th spin, defined as $\sigma_i^\pm = (\sigma_i^x \pm i\sigma_i^y)/2$. $c_{i}^{\dagger}$ ($c_{i}$) is the fermionic creation (annihilation) operator satisfying $\{c_i,c_j^\dagger\}=\delta_{ij}$. Transforming to momentum space under periodic boundary conditions, with $c_i = \sum_k e^{i k x_i} c_k/\sqrt{N}$ and $k = 2\pi m/N$ ($m = -N/2+1, \dots, N/2$), the Hamiltonian becomes
\begin{align}
H_{\text{XY}}'=&-\sum_k \big\{i(J_x - J_y)\sin k (c_k^\dagger c_{-k}^\dagger + c_k c_{-k} )
- 2[(J_x + J_y)\cos k +h ]c_k^\dagger c_k \big\}  + hN.
\end{align}
After performing the Bogoliubov transformation \cite{Sachdev2011}, we can diagonalize the Hamiltonian as
\begin{equation}
H_{\text{XY}}^{'}=\sum_{k}E_{k}^{\rm XY}\gamma_{k}^{\dagger}\gamma_{k}+E_0^{\rm XY},
\end{equation}
in which $\gamma_k^\dagger$ ($\gamma_k$) represent the Bogoliubov quasiparticle creation (annihilation) operators. The corresponding energy spectrum and ground-state energy are
\begin{align}
\begin{split}
E_{k}^{\text{XY}}\!\!&=\!\!2\sqrt{[h+(J_{x}+J_{y})\cos k]^{2}+[(J_{x}-J_{y})\sin k]^{2}},  \\
E_0^{\text{XY}}\!\!&=\!\!-\frac{1}{2}\sum_k{E_{k}^{\text{XY}}},
\end{split}
\end{align}
respectively.

In the thermodynamic limit, i.e., $N\rightarrow\infty$, the energy density of the ground state becomes
\begin{align}
e_g^{\text{XY}} &= \lim_{N \to \infty} \frac{E_0^{\text{XY}}}{N}
= -\frac{1}{2\pi} \int_{0}^{\pi} E_k^{\text{XY}} dk.
\end{align}
The critical points where the QPT occurs can be identified by examining the closure of the energy gap, i.e., $\min_k (E_k^{XY}) = 0$. This leads to the conditions, i.e.,
\begin{equation}\label{eq:QPTc}
\begin{split}
h+(J_{x}+J_{y})\cos k&=0, \\
(J_{x}-J_{y})\sin k&=0,
\end{split}
\end{equation}
which should be satisfied simultaneously. This analysis reveals several important cases. For the isotropic XY model, i.e., $J_x = J_y = J$, the condition (\ref{eq:QPTc}) is simplified as
\begin{equation}
h+2J\cos k=0.
\end{equation}
The critical points occur at $h_c=\pm2J$, as $\cos k\in[-1,1]$, where the system enters the fully $z$-polarized paramagnetic phase. This symmetric behavior reflects the rotational invariance of the isotropic model. 
In the general case with anisotropic couplings, i.e., $J_x \neq J_y$, the condition (\ref{eq:QPTc}) requires $\sin k=0$, namely, $k=0$ or $k=\pi$, leading to the critical fields as
\begin{align}
h_c &= -(J_x + J_y), \quad \text{for $k=0$}, \nonumber \\
h_c &= J_x + J_y, \quad \text{for $k=\pi$}.
\label{eq11}
\end{align}
The critical behavior thus depends on the sum of the two coupling constants. 
The TFI model is a important special case, i.e., $J_y=0$ and $J_x=J>0$,  which yields
\begin{align}
h_c &= -J, \quad \text{for $k=0$},  \nonumber \\
h_c &= J, \quad \text{for $k=\pi$}.
\end{align}
For simplicity, we adopt the convention of a positive transverse field $h > 0$ throughout this work. Under this convention, the QPT driven by increasing $h$ from zero occurs at the critical field $h_c = J_x + J_y$, which corresponds to the gap closing at momentum $k = \pi$ as given in Equation~\eqref{eq11}. This marks the transition from a ferromagnetically ordered phase to a paramagnetic phase. The alternative critical point $h_c = -(J_x+J_y)$ for $k=0$ corresponds to a negative transverse field; it is physically equivalent to the positive-field case under a global spin rotation (e.g., a $\pi$ rotation around the $x$-axis flips the sign of $h$).

\subsection{Network of $N$-DOPOs}

\begin{figure}
\centering
\includegraphics[width=12cm]{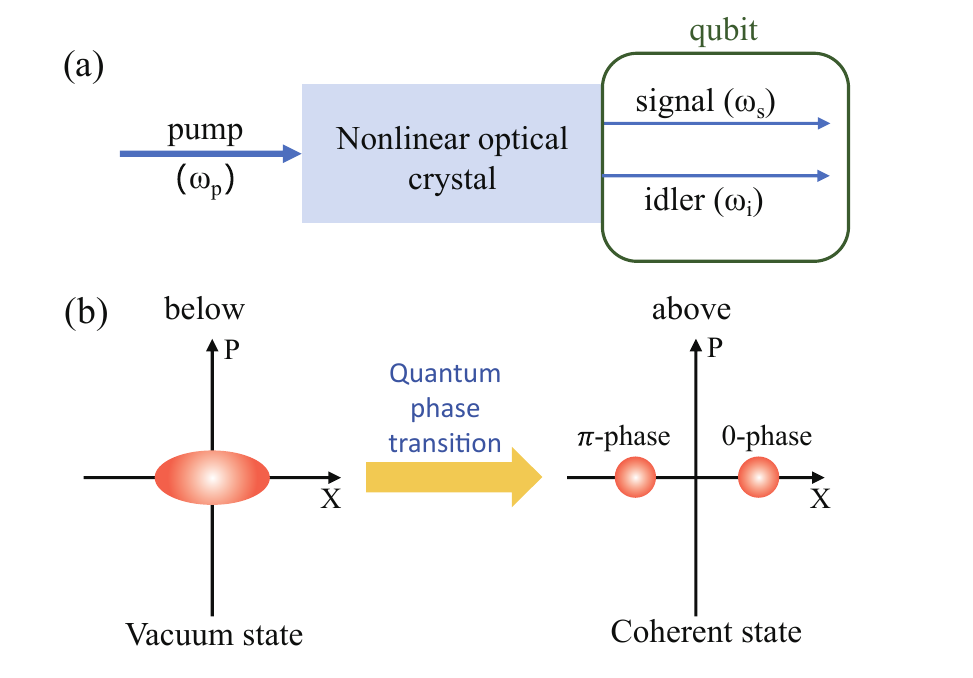}
\caption{(a) Schematic representation of a DOPO, a key component for constructing a CIM. (b) Simplified phase-space illustration of a DOPO showing the vacuum state below the threshold and the two coherent states with phases $0$ and $\pi$ above the threshold.}
\label{fig2}
\end{figure}

In a OPO with second-order nonlinear effects, a pump field at frequency
$\omega_p$ undergoes parametric down-conversion within the nonlinear medium, yielding a signal field with frequency $\omega_s$ and an idler field with frequency $\omega_i$. In the degenerate regime considered in this work, the signal and idler frequencies are identical, i.e., $\omega_s=\omega_i=\omega_p/2$. As illustrated in Figure~\ref{fig2}~(a), the Hamiltonian of the system can be expressed as
\begin{align}
H_{\text{D}} = H_0 + H_{\text{int}},
\end{align}
where
\begin{align}
H_0&= \omega_s a_s^\dagger a_s +  \omega_p a_p^\dagger a_p, \\
H_{\text{int}}&=i\frac{\kappa}{2} a_s^{\dagger 2} a_{p}  +i\varepsilon a^\dagger_p e^{-i\omega_p t}+\textrm{H.c.}
\end{align}
Here $a_s^\dagger$ ($a_s$) and $a_p^\dagger$ ($a_p$) are the creation (annihilation) operators for the signal and pump fields, respectively. The coupling constant $\kappa$ quantifies the strength of the second-order nonlinear interaction, and $\varepsilon$ is the amplitude of the external coherent drive applied to the pump mode.
When there is a detuning $\Delta = \omega_s - (\omega_p/2)$ between the signal and the pump fields, with respect to the free Hamiltonian
\begin{equation}
H_{f} = \frac{\omega_p}{2} a_s^\dagger a_s +  \omega_p a_p^\dagger a_p,
\end{equation}
the Hamiltonians are transformed to the interaction picture as
\begin{align}
H_{0}' = & \Delta a_s^\dagger a_s,  \\
H_{\text{int}}' = &i\frac{\kappa}{2}  a_s^{\dagger 2} a_{p} 
+ i\varepsilon a^\dagger_p+\textrm{H.c.}
\end{align}
Following the Heisenberg-Langevin approach \cite{Scully1997}, the temporal evolution of the pump field is described by
\begin{align}
\frac{da_p}{dt} = - \gamma_p a_p+\varepsilon - \frac{\kappa}{2} a_s^2, \label{eq:dp}\\
\end{align}
where $\gamma_p$ represents the decay rate of the pump mode. 
Assuming the pump field reaches its steady state rapidly, we adiabatically eliminate it to obtain an effective Hamiltonian for the signal mode as
\begin{equation}
H_{\text{D}} = \Delta a^\dagger a + i\frac{\kappa\varepsilon}{2\gamma_p}a^{\dagger2}+\textrm{H.c.}
\label{eqHcim}
\end{equation}
Hereafter, we omit the subscript $s$ for the signal field for simplicity.

Extending this description to $N$ nearest-neighbor-coupled DOPOs yields the generalized Hamiltonian
\begin{align}
    H_{\text{ND}} = & \sum_{i=1}^{N} \Delta a_i^\dagger a_i + \sum_{i=1}^{N} \left(i\frac{\kappa\varepsilon}{2\gamma_p} a_i^{\dagger 2}- J  a_i^\dagger a_{i+1}\right) + \textrm{H.c.},
     \label{eq:original_hamiltonian}
\end{align}
where $a_i^\dagger$ ($a_i$) is the creation (annihilation) operator for the $i$th DOPO, and $J$ quantifies the coupling strength between two adjacent DOPOs.
To diagonalize $H_{\text{ND}}$, we first move to momentum space via a Fourier transformation, which brings the Hamiltonian into the form
\begin{equation}
    H^{'}_{\text{ND}} = \sum_{k} \left( \epsilon_k a_k^\dagger a_k + i\frac{\kappa\varepsilon}{2\gamma_p} a_k^\dagger a_{-k}^\dagger +\textrm{H.c.}\right),\label{eq:Hk}
\end{equation}
where we have defined $\epsilon_k = \Delta - 2J \cos k$ for brevity.
A subsequent Bogoliubov transformation
\begin{equation}
    b_k = a_k\cosh r_k + a_{-k}^\dagger e^{i\theta} \sinh r_k
\end{equation}
is then applied, where the parameters are chosen as $\tanh(2r_k) = |\kappa\varepsilon| / (\gamma_p \epsilon_k)$ and $\theta = \arg(\kappa\varepsilon) + \pi/2$ to remove off-diagonal terms.
The resulting diagonal Hamiltonian is
\begin{equation}
    H^{'}_{\text{ND}} = \sum_{k} \Omega_k b_k^\dagger b_k + \frac{1}{2} \sum_{k} \left( \Omega_k - \epsilon_k \right),\label{eq:Hfinal}
\end{equation}
where the quasiparticle energy spectrum reads
\begin{equation}
    \Omega_k = \sqrt{\epsilon_k^2 - \left|\frac{\kappa \varepsilon}{\gamma_p}\right|^2},
\end{equation}
and the second sum corresponds to the zero-point energy contribution.

In this framework, the DOPO network exhibits a QPT from a normal phase (vacuum state) to a symmetry-broken phase (coherent state) when the pump intensity exceeds the threshold, as shown in Figure~\ref{fig2}(b). 
Although the intrinsic threshold transition of a DOPO involves a $\mathbb{Z}_2$ symmetry breaking that is mathematically reminiscent of magnetic ordering, it is distinct from the XY-model quantum criticality considered below. In our scheme, the CIM is operated in its above-threshold regime, while the critical behavior of the XY model is encoded through the engineered spectrum of the DOPO network.

\subsection{Energy-Spectrum Matching}

We now present a rigorous mapping between the energy spectra of the XY model and that of the $N$-DOPO network. Defining $J_s = J_x + J_y$, $J_d = J_x - J_y$, and $D = \left|\kappa\varepsilon/\gamma_p\right|$, the squared energy spectrum of the XY model is given by
\begin{align}
(E_{k}^{\text{XY}})^2 = 4(h^2 + J_d^2) + 8J_sh\cos k + 16J_xJ_y\cos^2 k .\label{eq:Ek}
\end{align}
Meanwhile, the squared energy spectrum of the DOPO network reads
\begin{align}
\Omega_k^2 = \Delta^2 - D^2 - 4\Delta J\cos k + 4J^2\cos^2 k .\label{eq:Omegak}
\end{align}
Requiring the two spectra to coincide for all $k$, we obtain the exact mapping
{\begin{equation}
\begin{split}
J      &= 2\sqrt{J_xJ_y},\\
\Delta &= -\frac{h J_s}{\sqrt{J_xJ_y}},\\
D^2    &= J_d^2\!\left(\frac{h^2}{J_xJ_y} - 4\right).
\end{split}
\label{eq:31}
\end{equation}}
Through this transformation, the critical behavior of the XY model can be faithfully studied in the DOPO network. For the anisotropic case $J_x\neq J_y$, it requires $D^2\geq 0$. For a positive transverse field $h>0$, this condition yields
\begin{equation}
h\geq h_{\min}=2\sqrt{J_xJ_y}.
\end{equation}
Therefore, the present DOPO mapping does not cover the regime from $h=0$ to $h_{\min}$. Instead, it provides an exact spectral realization in an experimentally-accessible regime from a field slightly-lower than the critical point to a strong field. Importantly, the quantum critical point always lies within the accessible domain because
\begin{equation}
h_c=J_x+J_y
\geq
2\sqrt{J_xJ_y}
=
h_{\min}.
\end{equation}
Thus, the closing of the energy gap can always be captured by the proposed DOPO network. For the near-TFI example adopted in the original manuscript, i.e., $J_x=1$ and $J_y=0.01$, one obtains $h_{\min}=0.2<h_c=1.01$, which shows that the critical point, together with two phases on its both sides, remains accessible. Nevertheless, the deep in the ferromagnetically ordered phase along $x$ direction, i.e., $0\leq h<h_{\min}$, can not be directly observed by the present mapping. In the thermodynamic limit, the energy density satisfies
\begin{equation}
e_g^{\mathrm{D}} = -\,e_g^{\mathrm{XY}}
                       + \frac{h J_s}{2\sqrt{J_xJ_y}}.
\end{equation}
This relationship implies that the non-analyticities in the ground-state energy of the XY model, which characterize the QPT, are directly inherited by the DOPO network, albeit with a linear shift and scaling.

We next clarify whether the parameter variation required by the spectral mapping can drive an intrinsic dissipative bifurcation of the DOPO network. In a realistic optical
system, the signal mode has a nonvanishing decay rate $\gamma_s$. 
For each $k$-mode in Equation~\eqref{eq:Hk}, the equation of motion can be written as
\begin{align}
\frac{d}{dt} \begin{pmatrix} a_k \\ a_{-k}^\dagger \end{pmatrix} = 
\begin{pmatrix} -\gamma_s - i \varepsilon_k & D \\ D & -\gamma_s + i \varepsilon_k \end{pmatrix} \begin{pmatrix} a_k \\ a_{-k}^\dagger \end{pmatrix}.
\end{align}
The eigenvalues of the drift matrix are
\begin{equation}
\lambda_{k,\pm}
=
-\gamma_s\pm\sqrt{D^2-\epsilon_k^2}.
\label{eq:drift_eigenvalues}
\end{equation}
Consequently, throughout the entire accessible mapping parameter regime, the drift eigenvalues can be written as $\lambda_{k,\pm}=-\gamma_s\pm i\Omega_k$ and hence $\rm{Re}\lambda_{k,\pm}=-\gamma_s<0$. Therefore, no drift eigenvalue crosses the imaginary axis and no additional steady-state branch emerges as the mapped parameters are varied. The DOPO network therefore remains within the same dynamically stable degenerate parametric regime. At the critical point of the mapped XY model, the $k=\pi$ mode satisfies
$\Omega_\pi=0$. Nevertheless, the corresponding drift eigenvalues remain
\begin{equation}
\lambda_{\pi,+}=\lambda_{\pi,-}=-\gamma_s<0,
\end{equation}
so the mapped gap-closing condition does not represent a dissipative instability of the
DOPO network. The actual parametric-oscillation threshold would instead be determined by
$\lambda_{k,+}=0$, which gives
\begin{equation}
D^2=\epsilon_k^2+\gamma_s^2.
\label{eq:true_dopo_threshold}
\end{equation}
This threshold is never reached along the mapped parameter regime because
\begin{equation}
\epsilon_k^2+\gamma_s^2-D^2
=
\Omega_k^2+\gamma_s^2
>0.
\label{eq:threshold_margin}
\end{equation}
In particular, even at $\Omega_\pi=0$, the system remains distinguished from the threshold of the DOPO by the nonvanishing $\gamma_s^2$. This situation is fundamentally different from the detuning-induced spectral phase transition reported for optical parametric oscillators in Ref.~\cite{Arkadev2021NC,Arkadev2023NP}. In that work, tuning the cavity phase detuning $\Delta\phi$ changes the oscillation selected by the OPO from a degenerate regime, in which the signal and idler overlap at the half-harmonic frequency $\omega_0$, to a non-degenerate regime with two distinct frequency branches $\omega_0\pm\delta\omega$. The square-root splitting of these branches therefore represents a transition between two spectral regimes of the OPO itself.

By contrast, the DOPOs considered in the present mapping remain in the degenerate parametric configuration throughout the mapped parameter trajectory. Varying the mapped detuning $\Delta$, together with the parametric amplitude $D$ according to Equation~\eqref{eq:31}, does not produce signal--idler frequency splitting, steady-state branch switching, or a transition into a non-degenerate OPO regime. Moreover, the momentum $k$ in our model labels a spatial normal mode of the coupled-DOPO network, rather than the optical signal--idler frequency offset $\delta\omega$ considered in Ref.~\cite{Arkadev2021NC}. Therefore, $\Omega_\pi=0$ describes the softening of the mapped $k=\pi$ network mode and not a degenerate-to-nondegenerate transition of the optical system.

Having established that the mapped parameter regime does not cross the dissipative bifurcation threshold, we next examine how finite loss modifies the experimentally
observable signature of the mapped XY critical point. Although the DOPO network remains within the same stable regime, the decay rate $\gamma_s$ changes the real-valued Bogoliubov frequencies into complex response poles and provides a finite linewidth. After performing a Fourier transformation, i.e., $a_k(t) = \int \frac{d\omega}{2\pi} a_k(\omega)e^{-i\omega t}$, the corresponding response function~\cite{Clerk2010RMP} is
\begin{align}
\chi_k(\omega) = \frac{1}{(\gamma_s - i\omega)^2 + \Omega_k^2} \begin{pmatrix} \gamma_s - i(\omega + \epsilon_k) & D \\ D & \gamma_s - i(\omega - \epsilon_k) \end{pmatrix}.
\end{align}
Thus, the ideal gap closing of the mapped XY model appears in the dissipative DOPO as a softening of the response pole, while the finite decay rate $\gamma_s$ provides a linewidth. In particular, at the critical momentum $k=\pi$ and at zero frequency, the soft-mode response can be characterized by
\begin{align}
R_\pi(0)\propto
\frac{1}{\gamma_s^2+\Omega_\pi^2}.
\end{align}
Therefore, the divergence expected in the closed XY model is rounded into a finite response peak in the dissipative DOPO system. The peak height and width are controlled by $\gamma_s$, while the peak position still identifies the mapped XY critical point through $\Omega_\pi=0$. 
In this case, a broadened spectral functional can be written as
\begin{equation}
\tilde{e}_g^{\mathrm{D}}= -\frac{1}{2\pi} \int_0^\pi \sqrt{\Omega_k^2 + \gamma_s^2} \, dk.
\end{equation}
Its second derivative provides a convenient measure of the dissipatively broadened susceptibility, rather than a thermodynamic ground-state susceptibility of the DOPO
system itself. The phase assignment is made by first using the mapping relation, i.e., Equation~\eqref{eq:31}, to reconstruct the effective transverse field of the corresponding XY model,
\begin{align}
h_{\rm eff}
=
-\Delta
\frac{\sqrt{J_xJ_y}}{J_x+J_y}.
\end{align}
Then $h_{\rm eff}$ is compared with the critical point $h_c$. For $h_{\rm eff}<h_c$, the mapped closed XY model lies on the ordered side. For $h_{\rm eff}>h_c$, it lies in the paramagnetic phase, and the critical point is identified by $h_{\rm eff}=h_c$, which coincides with the soft-mode condition $\Omega_\pi=0$.

\section{Results and Discussion}
\label{sec:results and discussion}

\begin{figure*}
\centering
\includegraphics[width=0.85\linewidth,clip=true, trim=0 10 5 5]{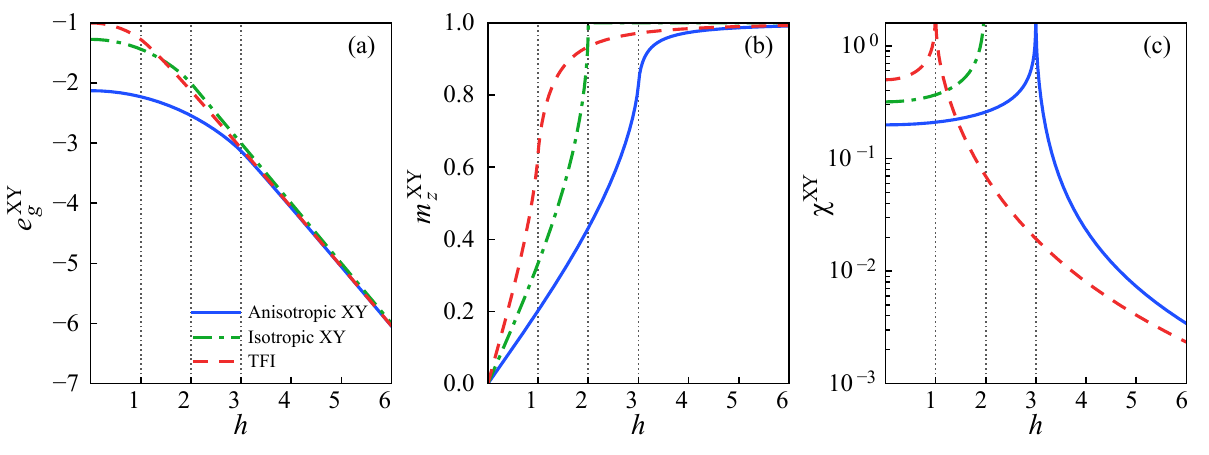}
\caption{Numerical results for three quantum-spin-chain models under an external field $h$: (a) Ground-state energy density $e_g^{\mathrm{XY}}$, (b) magnetization $m_z^{\mathrm{XY}} = -\partial e_g^{\mathrm{XY}}/\partial h$, and (c) magnetic susceptibility $\chi^{\mathrm{XY}} = -\partial^2 e_g^{\mathrm{XY}}/\partial h^2$. Note that sub-figure (c) uses a logarithmic scale.
The curves correspond to the anisotropic XY model (\(J_x = 2, J_y = 1\), blue solid line), the isotropic XY model (\(J_x = J_y = 1\), green dash-dotted line), and the transverse-field Ising model (\(J_x = 1, J_y = 0\), red dashed line). The vertical dotted lines mark the calculated critical fields $h_c$ for each model.}
\label{fig1}
\end{figure*}

As illustrated in Figure~\ref{fig1}, we consider three quantum spin-chain models: the anisotropic XY model with parameters \(J_x=2\) and \(J_y=1\), the isotropic XY model with \(J_x=J_y=1\), and the TFI model with \(J_x=1\) and \(J_y=0\). 
The anisotropic XY and TFI models exhibit Ising-type second-order QPT driven by the transverse field. The isotropic XY model represents the \(U(1)\)-symmetric limit, where the spectrum is gapless for \(|h|\le 2J\) and the positive-field boundary \(h_c=2J\) marks the transition into the fully polarized paramagnetic phase. 
In the thermodynamic limit, the TFI model and the anisotropic XY model with \(J_x>J_y\) possess a symmetry-broken ordered phase at small \(h\), characterized by a nonzero \(x\)-direction order parameter \(m_x\). In contrast, the isotropic XY model has a \(U(1)\) spin-rotation symmetry about the \(z\) axis and forms a gapless critical phase for \(|h|<2J\), with \(m_z=0\) at \(h=0\). As \(h\) increases from zero, the ground-state energy density $e_g^{\mathrm{XY}}$ and the longitudinal magnetization $m_z^{\mathrm{XY}}$ evolve continuously. At the critical field strength \(h_c = J_x + J_y\), each system undergoes a QPT into a paramagnetic phase, where the spins become predominantly aligned along the field direction. For \(h > h_c\), the longitudinal magnetization \(m_z\) increases monotonically toward its saturation value of \(+1\), whereas the transverse components \(m_x\) and \(m_y\) are suppressed to zero. The magnetic susceptibility \(\chi^{\mathrm{XY}}\) displays a divergence near \(h_c\), directly signaling the critical point. For the isotropic XY model in the paramagnetic phase, the first derivative of the ground-state energy density remains constant, leading to a vanishing second derivative. This accounts for the disappearance of its corresponding curve beyond \(h_c\) in the logarithmic-scale plot of Figure~\ref{fig1}(c). Collectively, this systematic analysis provides a clear physical interpretation and numerical verification for the simulation of quantum critical phenomena using a CIM.

\begin{figure*}[!htb]
\centering
\includegraphics[width=0.85\linewidth,clip=true, trim=0 20 5 20]{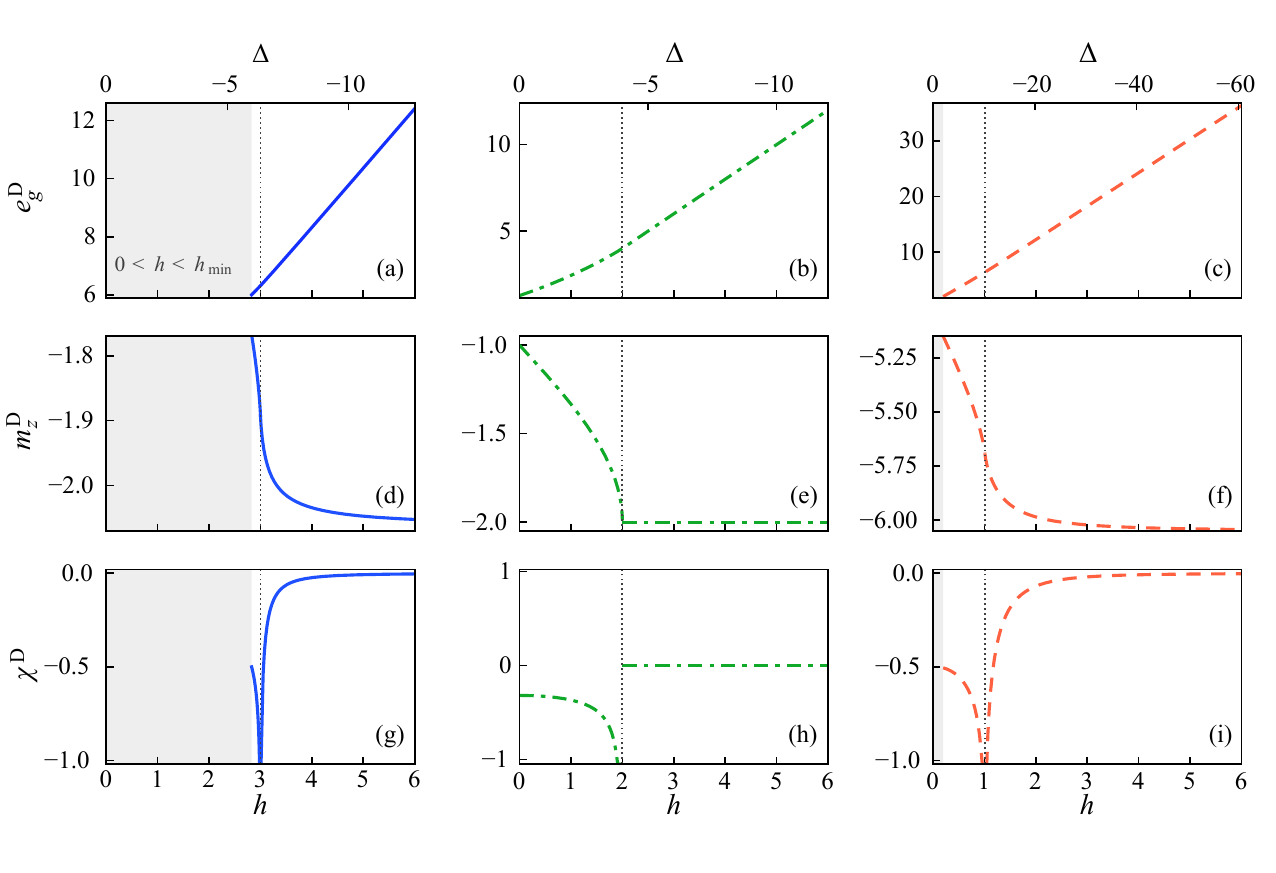}
\caption{The spectral signatures of the mapped XY critical points in the representation of the DOPO network. Columns show DOPO parameter regimes: (left) $J=2\sqrt{2}$, $\Delta = -3h/\sqrt{2}$; (middle) $J=2$, $\Delta = -2h$; (right) $J = 0.2$, $\Delta = -10.1h$. These correspond to anisotropic XY, isotropic XY, and near-TFI spin chains, respectively. From top to bottom, the panels display: (a--c) the ground-state energy density $e_g^{\mathrm{D}}$; (d--f) the longitudinal magnetization $m_z^{\mathrm{D}}$; and (g--i) the magnetic susceptibility $\chi^{\mathrm{D}}$.
Each panel features dual horizontal axes: the upper axis shows the DOPO detuning parameter $\Delta$, and the lower axis shows the corresponding transverse field $h$ in the XY model. 
Vertical dashed lines indicate the critical points at $\Delta_c = -2J - D$. In addition, the gray-shaded regions denote the field interval $0<h<h_{\min}=2\sqrt{J_xJ_y}$, where $D^2<0$ and the present finite-parameter DOPO mapping admits no real parametric-drive amplitude. The shading therefore marks an optically inaccessible region of the mapping, rather than an absent phase of the XY model. For the isotropic case $J_x=J_y$, $D=0$ and no such restriction arises.}
\label{fig4}
\end{figure*}

The simulated ground-state properties of the DOPO network are presented in Figure~\ref{fig4}. The three columns correspond to different regimes of the XY model mapped onto the optical system. The left column, with parameters $J=2\sqrt{2}$ and $\Delta=-3h/\sqrt{2}$, represents the anisotropic XY model. The middle column, with $J=2$, $\Delta=-2h$, and $D=0$, corresponds to the isotropic XY model. The right column approximates the TFI limit using a strongly anisotropic XY model with $J_y \ll J_x$. For concreteness, taking $J_x=1$ and $J_y=0.01$, the mapping in Equation~\eqref{eq:31} gives $J = 0.2$ and $\Delta = -10.1h$. 
For finite $J_y>0$, the mapping guarantees an exact equality $\Omega_k=E_k^{XY}$ for the corresponding anisotropic XY chain. Therefore, in the anisotropic and isotropic cases the gap closing of the XY model is exactly mapped onto the spectrum of the DOPO. In the right column, $J_y=0.01$ gives a realization of the quasi-TFI, whose convergence toward the strict TFI spectrum is analyzed below in Figure~\ref{fig:Energy-spectrum}. The vertical dashed line in the figure marks $\Delta_c = -2J - D$, which corresponds precisely to this critical point and manifests as a divergence of the susceptibility $\chi^{\mathrm{D}}$. It is worth noting that, under the mapping, both $\Delta$ and $D$ vary together with $h$, and the system always satisfies $\epsilon_k^2 \ge D^2$. Even at $\Delta_c$, where the gap closes, the system merely touches the phase boundary. 
The middle column, corresponding to the isotropic limit $J_x=J_y$, is a limiting case of the mapping with $D=0$. In this limit the parametric-amplification term vanishes,
and the mapped spectrum is realized by the linear detuned oscillator network. These results demonstrate that the critical point of the XY model can be located by measuring the susceptibility $\chi^{\mathrm{D}}$ of the DOPO network, thereby providing a feasible scheme for experimental studies of the QPT in the XY model on a CIM platform.

Because the strict TFI point $J_y=0$ is singular in the finite-parameter mapping of the DOPO, the right column of Figure~\ref{fig4} should be understood as a quasi-TFI realization rather than as an exact finite-parameter TFI mapping. 

To quantify this approximation, we perform a convergence analysis in Figure~\ref{fig:Energy-spectrum}. 
As shown in  Figure~\ref{fig:Energy-spectrum}(a), the spectral mapping between the DOPO network and the corresponding XY chain is almost exact. The issue is therefore not the validity of the mapping at $J_y=0.01$, but rather the convergence of this strongly-anisotropic XY model toward the exact TFI spectrum with $J_y=0$. To show this convergence explicit, we introduce $\eta=J_y/J_x$ and demonstrate that the parameter used in Figure~\ref{fig:Energy-spectrum}(b) satisfies
\begin{equation}
\lambda_{\mathrm{opt}}
=
\frac{4\eta}{(1+\eta)^2}
\end{equation}
with $\lambda_{\mathrm{opt}}=2J/|\Delta_c|$, at the critical point. Therefore, decreasing $\lambda_{\mathrm{opt}}$ in Figure~\ref{fig:Energy-spectrum}(b) is equivalent to a systematic reduction of $\eta$. The progressive convergence of the DOPO spectra toward the exact TFI spectrum near $k=\pi$ and $k=0$ demonstrates that the DOPO realization of the quasi-TFI continuously approaches the exact TFI spectrum as $\eta\to0$. From an experimental perspective, the quasi-TFI regime should be understood as a finite-anisotropy approximation rather than the singular limit $\eta\to0$. Therefore, approaching the strict TFI limit increases the required dynamic range between the detuning and parametric drive on the one hand and the inter-DOPO coupling on the other. In practice, the achievable tuning range and the finite linewidth $\gamma_s$ of the signal mode set a lower bound on how small $\eta$ can be made. Therefore, the smallest values of $\lambda_{\mathrm{opt}}$ shown in Figure~\ref{fig:Energy-spectrum}(b) are intended to illustrate the asymptotic convergence toward the TFI limit, rather than to prescribe experimentally-required operating points. Consequently, a finite anisotropy, such as $\eta=0.01$, provides a controlled quasi-TFI realization without requiring access to the singular optical limit. Recent CIM experiments have demonstrated programmable coupling, pump control, and long-term cavity stabilization, providing an experimental basis for exploring such finite quasi-TFI regimes~\cite{science2016Inagaki,science2016McMahon,Wei2026Light}.
\begin{figure*}[!hbtp]
    \centering
	\includegraphics[width=0.85\linewidth,clip=true, trim=10 200 30 80]{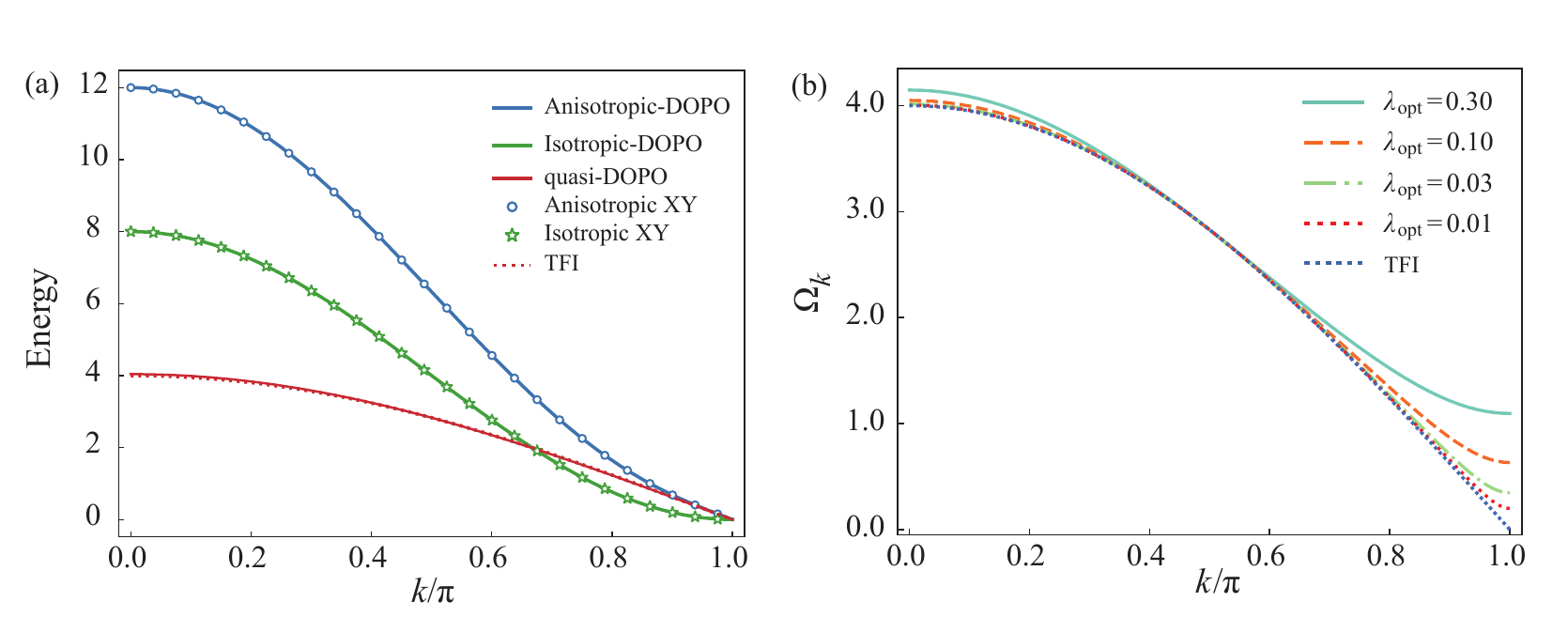}
\caption{Energy-spectrum correspondence between the DOPO network and the XY-family spin models. (a) Exact spectral matching between the DOPO quasiparticle spectra and the corresponding anisotropic XY, isotropic XY, and TFI spin-chain spectra at the critical point, i.e., $h=J_x+J_y$. For the anisotropic XY model, the parameters are $J_x=2$, $J_y=1$, while the corresponding parameters for the DOPO are $J=2\sqrt{2}$, $\Delta = -9/\sqrt{2}$ and $D^2=0.45$. For the isotropic XY model, the parameters are $J_x=1$ and $J_y=1$, while the corresponding parameters for the DOPO are $J=2$, $\Delta = -6$ and $D^2=0$. For the TFI model, the parameters are $J_x=1$ and $J_y=0$, while the corresponding parameters for the DOPO are $J = 0.2$ ($J_y=0.01$), $\Delta = -10.1$ and $D^2=96.06$. For the first two cases, the mapping $\Omega_k=E_k^{\mathrm{XY}}$ can be exactly fulfilled. For the TFI model, the energy spectrum of the DOPO is quite close to that of the TFI model.
(b) The convergence of the DOPO spectrum toward the TFI spectrum as $\lambda_{\rm opt}\rightarrow0$. The intrinsic optical ratio $\lambda_{\mathrm{opt}}=2J/|\Delta_c|$ is related to the XY
anisotropy ratio by $\lambda_{\mathrm{opt}}=4\eta/(1+\eta)^2$, where $\eta=J_y/J_x$. Hence, decreasing $\lambda_{\mathrm{opt}}$ from $0.30$ to $0.01$ corresponds to
a systematic reduction of $J_y/J_x$ and thus to an approach toward the TFI limit. The progressive suppression of the deviation near $k=\pi$ and $k=0$ demonstrates the convergence of the quasi-TFI DOPO realization of the exact TFI spectrum.
}
\label{fig:Energy-spectrum}
\end{figure*}
 
\begin{figure*}[!hbtp]
    \centering
	\includegraphics[width=0.85\linewidth,clip=true, trim=0 0 0 0]{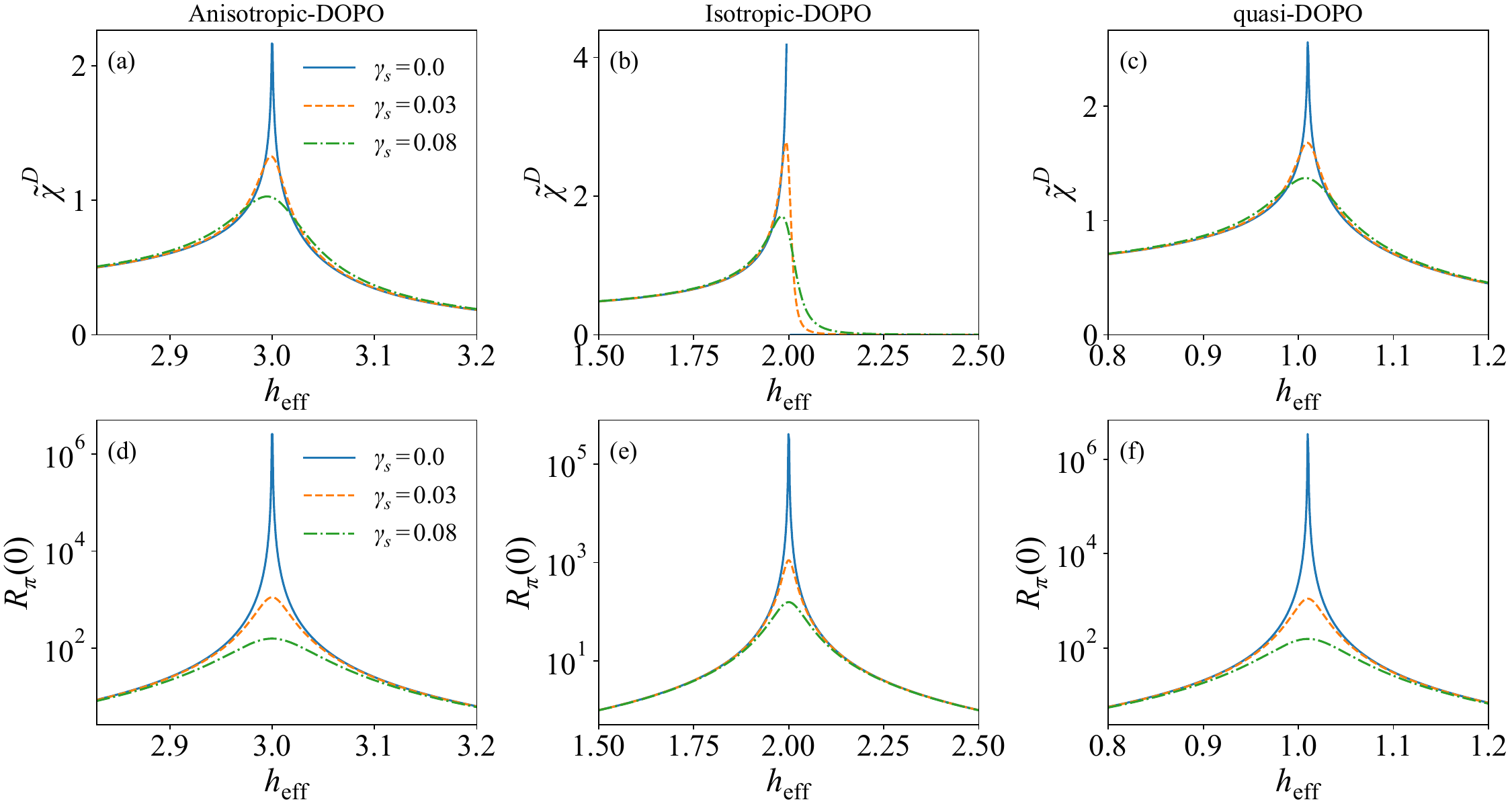}
\caption{Dissipative-DOPO signatures of quantum critical behavior in the mapped XY-family spin models.
Panels (a)--(c) show the broadened susceptibility $\tilde{\chi}^{D}$, i.e., the second derivative of the rescaled energy $\tilde{e}_g^D$ of the ground state, for the anisotropic-XY, isotropic-XY, and quasi-TFI mappings, respectively. Panels (d)--(f) show the corresponding obscured-mode response $R_\pi(0)$ at $k=\pi$ . The horizontal axis is the effective transverse field $h_{\rm eff}$, reconstructed from the detuning of the DOPO through the spectral mapping. Different line styles denote different dissipation rates $\gamma_s$ of the signal mode, and the vertical dotted lines mark the critical field $h_c$ of the corresponding XY model. 
As $h_{\rm eff}$ approaches $h_c$, both $\tilde{\chi}^{D}$ and $R_\pi(0)$ develop distinct peaks, indicating that the dissipative linear response of the DOPO can locate the critical point of the mapped XY model. For $\gamma_s=0$, $R_\pi(0)$ formally diverges at $\Omega_\pi=0$, whereas a nonvanishing $\gamma_s$ broadens this ideal singularity into a finite-height response peak.
}
\label{fig:dissipative}
\end{figure*}

To further clarify the role of dissipation, we now analyze the dissipative response of the mapped DOPO network. 
Figure~\ref{fig:dissipative} shows the dissipative-DOPO signatures of the mapped XY-family spin models. 
As shown in Figures~\ref{fig:dissipative}(a)--(c), the broadened susceptibility $\tilde{\chi}^D$ exhibits a clear peak when $h_{\rm eff}$ approaches $h_c$.  
The same critical point is also identified by the response in Figures~\ref{fig:dissipative}(d)-(f). At the critical point of the mapped XY, the $k=\pi$ mode satisfies $\Omega_\pi=0$, leading to an enhanced response. For $\gamma_s=0$, this expression formally diverges at $h_{\rm eff}=h_c$. In the plotted curves, this ideal divergence is represented by the sharpest peak, while finite $\gamma_s$ broadens it into a finite-height response peak.

The comparison between the upper and lower rows of Figure~\ref{fig:dissipative} shows that $\tilde{\chi}^D$ and $R_\pi(0)$ locate the same critical field for all three mappings.
Increasing $\gamma_s$ broadens the anomaly and suppresses the peak height, but it does not shift the position of the critical response. Therefore, the dissipative DOPO network
retains the ability to identify the critical point of the corresponding closed XY model, even though the ideal gap-closing singularity is replaced by a finite-linewidth optical
response. 
Based on this local critical response, 
\begin{figure*}[hbtp]
    \centering
	\includegraphics[width=0.85\linewidth,clip=true, trim=50 80 50 50]{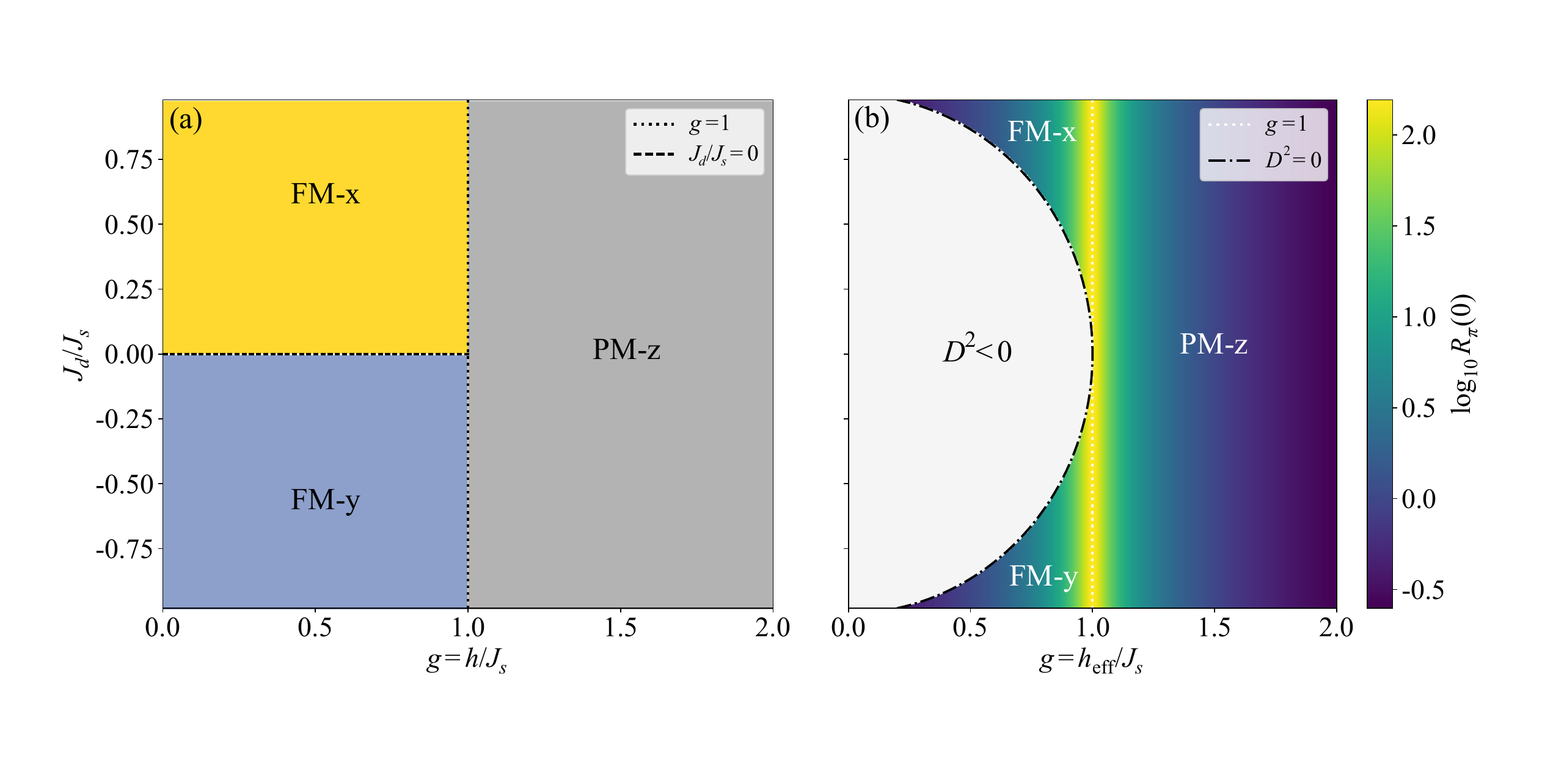}
\caption{
Phase diagram of the XY model and its identification through a dissipative response of the DOPO network. (a) Phase diagram of the XY model in the parameter space $(g,J_d/J_s)$, where $g=h/J_s$. The regions $g<1,J_d/J_s>0$ and $g<1,J_d/J_s<0$ correspond to the FM-$x$ and FM-$y$ phases, respectively, while $g>1$ corresponds to the
PM-$z$ phase. The dashed line denotes the critical line of the isotropic XY model, i.e., $J_d/J_s=0$, and the dotted line marks the critical line, i.e.,  $g=1$ or equivalently $h=h_c=J_s$. (b) Dissipative response map of the DOPO network, $\log_{10}R_\pi(0)$, expressed in the same normalized parameter space with $g=h_{\rm eff}/J_s$. The bright vertical ridge centered at $g=1$ represents the finite-linewidth enhancement of the $k=\pi$ mode and therefore locates the critical line of the mapped XY model. Combining the position relative to this ridge with the sign of $J_d/J_s$ allows one to identify the corresponding phases of the XY model within the accessible region $D^2\ge0$. The black dash-dotted curve denotes the boundary with $D^2=0$, and the shaded region with $D^2<0$ is inaccessible to the present finite-parameter mapping of the DOPO network. Importantly, panel (b) is an optical response map rather than a thermodynamic phase diagram of the DOPO network itself.
}
\label{fig:diagram}
\end{figure*}
Figure~\ref{fig:diagram} further illustrates how the dissipative response of the DOPO can be used to identify the phases of the XY model. Figure~\ref{fig:diagram}(a) shows the phase diagram of the XY model in the normalized parameter space $(g,J_d/J_s)$, where $g=h/J_s$. In the ordered regime $g<1$, the ferromagnetic ordering direction is selected by the sign of $J_d/J_s$. Positive $J_d/J_s$ leads to the FM-$x$ phase, while negative $J_d/J_s$ leads to the FM-$y$ phase. Once the normalized transverse field crosses the critical line $g=1$, the system becomes $z$-polarized and enters the paramagnetic phase PM-$z$. The line $J_d/J_s=0$ corresponds to the isotropic XY case. Figure~\ref{fig:diagram}(b) in contrast, is not a phase diagram of the optical system. It is a dissipative response map of the DOPO network, $\log_{10}R_\pi(0)$, in the same normalized parameter space, with the horizontal coordinate reconstructed from the detuning of the DOPO network as $g=h_{\rm eff}/J_s$. The bright vertical ridge at $g=1$ corresponds to the enhancement of the $k=\pi$ mode and therefore locates the critical line of the corresponding XY model. The black dash-dotted curve denotes the $D^2=0$ boundary, and the shaded region with $D^2<0$ is inaccessible to the present finite-parameter mapping of the DOPO network. 
Within the optically accessible region $D^2\ge 0$, combining the position of the broadened soft-mode ridge with the sign of $J_d/J_s$ allows one to infer the
corresponding phase of the mapped XY model, i.e., the region $g<1,J_d/J_s>0$ corresponds to FM-$x$, the region $g<1,J_d/J_s<0$ corresponds to FM-$y$, and $g>1$ corresponds to PM-$z$. Therefore, the dissipative DOPO network acts as a spectroscopic readout of the phase diagram of the XY. Figure~\ref{fig:diagram}(b) is not a thermodynamic phase diagram of the DOPO network itself, but rather an optical response map from which the phases of the corresponding XY model can be inferred.

\section{Conclusion}
\label{sec:conclusion}

In this work, we established a spectral mapping between aone-dimensional XY spin chain and a network of coupled DOPOs. For finite $J_x>0$ and $J_y>0$, the mapping reproduces the spectrum of the XY model exactly and therefore encodes the gap closing at the quantum critical point in experimentally tunable optical parameters. We emphasize that the present construction has a well-defined domain of validity. For anisotropic couplings $J_x\neq J_y$, the mapped parametric drive is real only for $h\geq2\sqrt{J_xJ_y}$. Hence, the scheme directly accesses the critical point and a near-critical strong-field window, but does not generally reproduce the complete evolution from zero field through the entire ordered phase. The strict transverse-field Ising limit $J_y=0$ is singular under the finite-parameter mapping. Nevertheless, a strongly anisotropic chain with $0<J_y\ll J_x$ provides a quasi-TFI regime. When finite loss of the signal mode is included, the ideal gap-closing singularity is rounded into a finite-linewidth response peak. The peak becomes broader and lower as $\gamma_s$ increases, while its center remains fixed by $\Omega_\pi=0$, allowing the dissipative response of the DOPO to serve as a spectroscopic readout of the critical line of the mapped XY.

Accordingly, the principal result of this work is not a finite-parameter realization of the QPT of the XY model in the DOPO network itself, but a tunable photonic
platform for spectrally emulating and detecting critical behavior of the corresponding XY family of the spin chains. The DOPO network remains in the same dynamically stable
dissipative regime, while its spectrum and linear response encode the criticality of the mapped spin model. This approach may be extended to other spin models and higher-dimensional photonic networks, where optical nonlinearity and dissipation can provide additional routes for probing critical phenomena.

\medskip
\textbf{Acknowledgments}

This work is supported by the National Natural Science Foundation of China under Grant No.~62461160263, and Quantum Science and Technology-National Science and Technology Major Project (2023ZD0300200), and Guangdong Provincial Quantum Science Strategic Initiative under Grant No.~GDZX2505004.

\bibliographystyle{MSP}

\end{document}